\documentclass{aa} 
\usepackage[varg]{txfonts}
\usepackage{graphicx}
\usepackage{latexsym}
\usepackage{natbib} 
\usepackage{amssymb,mathabx}
\bibpunct{(}{)}{;}{a}{}{,}
\begin{document}
\title{Short-term variability and mass loss in Be stars
}
\subtitle{II. Physical taxonomy of photometric variability\\ observed by the Kepler spacecraft}
\titlerunning{Short-term variability and mass loss in Be stars II}
\author{Th.~Rivinius\inst{1} \and D.~Baade\inst{2} \and A.~C.~Carciofi\inst{3} 
}
%
%
%
\institute{ESO --- European Organisation for Astronomical Research in the
  Southern Hemisphere, Casilla 19001, Santiago 19, Chile,
  \email{triviniu@eso.org} \and
ESO --- European Organisation for Astronomical Research in the Southern
Hemisphere, Karl-Schwarzschild-Str. 2, 85748 Garching, Germany,
\email{dbaade@eso.org}
\and 
Instituto de Astronomia, Geof\'isica e Ci\^encias Atmosf\'ericas, Universidade
de S\~ao Paulo, 05508-900, S\~ao Paulo, SP, Brazil
%
%
}
\date{Received: $<$date$>$; accepted: $<$date$>$}
\abstract%
{Classical Be stars have been established as pulsating stars. Space-based
  photometric monitoring missions contributed significantly to that result.
  However, whether Be stars are just rapidly rotating SPB or
  $\mathrm{\beta}$~Cep stars, or whether they have to be understood
  differently, remains debated in the view of their highly complex power
  spectra.}
{Kepler data of three known Be stars are re-visited to establish their
  pulsational nature and assess the properties of additional, non-pulsational
  variations.  The three program stars turned out to be one inactive Be star,
  one active, continuously outbursting Be star, and one Be star transiting
  from a non-outbursting into an outbursting phase, thus forming an excellent
  sample to distill properties of Be stars in the various phases of their
  life-cycle.  }
{The Kepler data was first cleaned from any long-term variability with
  Lomb-Scargle based pre-whitening. Then a Lomb-Scargle analysis of the
  remaining short-term variations was compared to a wavelet analysis of the
  cleaned data. This offers a new view on the variability, as it enables us to
  see the temporal evolution of the variability and phase relations between
  supposed beating phenomena, which are typically not visualized in a
  Lomb-Scargle analysis.}
{The short-term photometric variability of Be stars must be disentangled into
  a stellar and a circumstellar part. The stellar part is on the whole not
  different from what is seen in non-Be stars. However, some of the observed
  phenomena might be to be due to resonant mode coupling, a mechanism not
  typically considered for B-type stars. Short-term circumstellar variability
  comes in the form of either a group of relatively well-defined, short-lived
  frequencies during outbursts, which are called \v{S}tefl frequencies, and
  broad bumps in the power spectra, indicating aperiodic variability on a
  time scale similar to typical low-order $g$-mode pulsation frequencies,
  rather than true periodicity.}
{From a stellar pulsation perspective, Be stars are rapidly rotating SPB
  stars, that is they pulsate in low order $g$-modes, even if the rapid rotation
  can project the observed frequencies into the traditional high-order
  $p$-mode regime above about 4\,c/d. However, when a circumstellar disk is
  present, Be star power spectra are complicated by both cyclic, or periodic, and
  aperiodic circumstellar phenomena, possibly even dominating the power
  spectrum.}
\maketitle
%

\section{Introduction}  
Classical Be stars are rapidly rotating B stars, surrounded by a self-ejected
gaseous disk, the evolution of which is governed by viscosity. As a class,
they are known to pulsate in non-radial modes, but not to harbor large scale
magnetic fields \citep[see][for a review]{2013A&ARv..21...69R}. In general, Be
stars and their disks could be considered as fairly well understood, was it
not for the one central question that is still open: How are these disks
formed? The physical mechanism ejecting matter from the stellar surface with
sufficient energy and angular momentum to form a Keplerian disk, sometimes
called the Be phenomenon, remains elusive.

There is evidence that this Be phenomenon, at least in some stars, is related
to non-radial pulsation. This has first been shown by means of spectroscopy in
\object{\mbox{$\mathrm{\mu}$}~Cen} \citep[see, e.g.,][]{1998cvsw.conf..207R},
but the observational cost of this finding was high; spectra were taken daily
for several months each year over a number of years. Similar data sets
obtained for other Be stars suggested similar relations, but not as firm as
one would have wished.

With the launch of spacecraft dedicated to obtain photometric time series of
stars and stellar fields in the visual domain, it has become possible to take
data sets combining high precision with high cadence and very long time base,
the same combination that has enabled the spectroscopic
break-through. Although these missions are usually designed to search for
extra-solar planets or take data for the purpose of asteroseismology, they
serve equally well for the analysis of the light variations of Be stars. Since
Be stars are instrinsically bright and numerous, almost any field of stars as
large as the Kepler one and not too far away from the Galactic plane will host
at least some of them.

This series of papers has been inspired by the analysis of Be star photometric
data taken with the BRITE Constellation satellites
\citep{2014PASP..126..573W}, and \citet[called Paper I from here on]{PaperI}
present some of the first science results obtained from BRITE Constellation
data. Paper I presents the analysis of the BRITE data of
\object{\mbox{$\mathrm{\eta}$}~Cen} and \mbox{$\mathrm{\mu}$}~Cen, and much of
the findings here are already seen in BRITE data, though due to the lower
precision of BRITE, which has 3\,cm aperture compared to the 95\,cm of the
telescope aboard Kepler { and a less radiation-hard design}, not
necessarily with the same confidence. In any case, it became clear that every
space-based photometry mission has a partly unique potential, that could, and
should, be combined with results from other photometry missions and ground
based observations to unravel the mystery of the Be phenomenon.

The unique potential for Be stars observed with the Kepler spacecraft
\citep{2010ApJ...713L..79K} is without doubt the time base { combined with
  precision}, four full years of near-continuous observations with a cadence
of about half an hour. Be stars benefit in particular because they are
variable on a wide range of time scales from hours (and sometimes even
minutes, but only in the very high energy domain) to decades. In the
literature three Be stars have been published that were observed during the
original Kepler mission \citep{2011MNRAS.413.2403B,2015MNRAS.450.3015K}.

In the following Sect.~\ref{sec:data}, these three stars are introduced and
the analysis method employed in this work is described, which goes somewhat
beyond the methods used in the literature by using not only Lomb-Scargle based
Fourier analysis, but as well wavelet analysis. In Sect.~\ref{sec:results} the
results are presented, which do not contradict, but certainly extend the
ones reported in the literature, and the implications and interpretations are
discussed in Sect.~\ref{sec:discussion}. Finally, the results are put into a
generalized picture in Sect.~\ref{sec:conclusion}, and an outlook for future
papers of this series is given.

\begin{figure*}[t]
\begin{center}
\includegraphics[angle=0,width=18cm,clip]{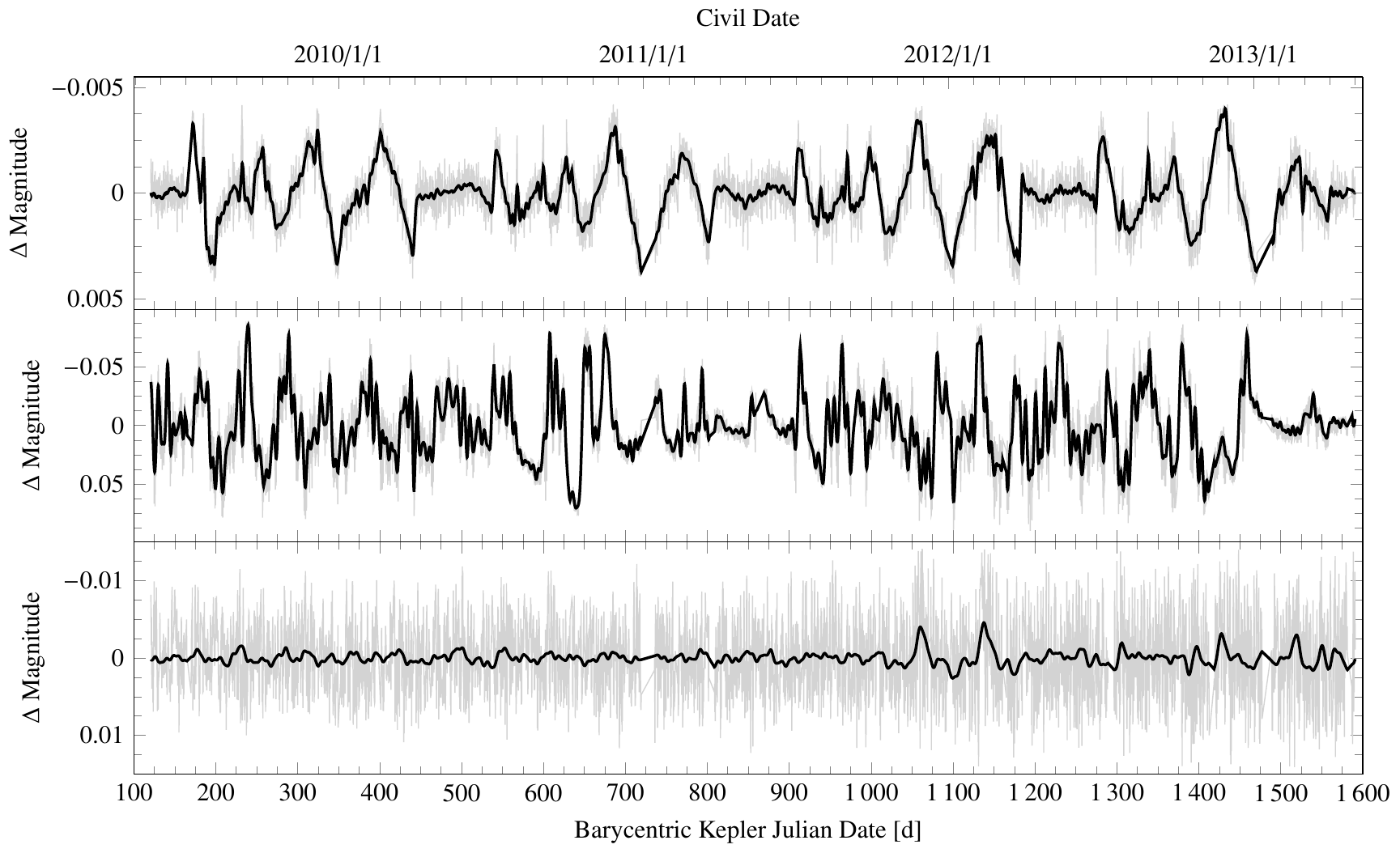}%

\end{center}
\caption[xx]{\label{fig:trends} Long-term part of the variability removed by
  high-pass filtering in black, original data gray in background. For this
  plot, linear trends have been removed beforehand. Shown are, from top to
  bottom, \mbox{ALS}10705, \mbox{StH$\alpha$}166, and \mbox{HD}186567. There
  is a difference in vertical scale, and hence amplitude of the variations.}
\end{figure*}

\section{Targets and analysis methods}  \label{sec:data}

Before introducing the targets, some terminology should be made clear.
Historically, an ``active Be star'' was typically understood to be one with
emission, but ``Be star activity'' was often identified with outbursts. This
is somewhat inconsistent, and since the understanding of Be stars has reached
a level at which this inconsistency becomes an obstacle, it should be
resolved. Here and in the following an inactive Be star is one without any
detectable disk, while an active Be star has such a disk. This designation is
observationally easy to probe, but makes no distinction between stars that
currently build a disk and stars in which the disk is decaying, so for the
former we introduce the term that it is a disk-feeding Be star. Not
withstanding the possibility that a star may build a disk in a continuous
manner, that is without outbursts, a star showing repeated and strong outbursts
is certainly in a feeding phase, { even though in the time span between two successive outbursts the star is not losing mass and the inner disk is technically in a dissipating phase}.

\subsection{\mbox{ALS}10705 (\object{KIC8057661})}
\mbox{ALS}10705 was found to show emission lines by
\citet{1999A&AS..134..255K}. From Kepler data \citet{2013A&A...557L..10N}
obtain 1.018\,d (0.982\,c/d) as dominant period, which they attribute to
rotation.  Two spectra were taken with ESPaDOnS at the CFHT in July 2010, a
bit more than a year after the beginning of the Kepler observations.  They do
not show any line emission. From these spectra, \citet{2011MNRAS.413.2403B}
obtain $v\sin i=49\pm5\,{\mathrm{km\,s}}^{-1}$, while from Str\"omgren
photometry they derive $T_{\mathrm{eff}}=21$\,kK and $\log g=4.2$. They give a
hybrid SPB and $\mathrm{\beta}$~Cep nature for the pulsational properties and
remark that \mbox{ALS}10705 should be considered as a currently inactive Be
star, seen at low inclination { of the rotational axis}.

\subsection{\mbox{StH$\alpha$}166 (\object{KIC6954726})}

\mbox{StH$\alpha$}166 was first noted to be an emission line star by
\citet{1986ApJ...300..779S} and confirmed to be a Be star by
\citet{1988AJ.....96..777D}. The Kepler Input Catalogue lists an effective
temperature of about 17\,kK and a $\log g$ of 4.8.

\citet{2009AcA....59...33P} identify the star as aperiodic variable in the
ASAS photometric database, while based on Kepler observations
\citet{2012AJ....143..101M} classify the variations as binary or rotational with
a dominant frequency of 1.03\,c/d, and \citet{2013A&A...560A...4R} derive a
rotation period of 44.5\,d (0.02\,c/d). \citet{2011MNRAS.413.2403B} note the
star has frequency groupings, but do not give a physical interpretation and
rather mention $g$-mode pulsation or rotation as possibilities.

\citeauthor{2011MNRAS.413.2403B} also derive $v\sin
i=160\,{\mathrm{km\,s}}^{-1}$, again based on two ESPaDOnS spectra, but do not
derive other parameters.  The spectra are somewhat noisy, but the ratio of
\ion{He}{i}\,4471 vs.\ \ion{Mg}{ii}\,4481 does not agree with a temperature as
cool as 17\,kK, and as well the presence of the \ion{Si}{iii}\,4553/68/73
triplet confirms that the star must rather be of earlier B subtype. Yet, the
absence of \ion{He}{ii} lines limits the $T_{\mathrm{eff}}$ to lower than
about 25\,kK.

The \mbox{H}$\mathrm{\alpha}$ emission line was symmetrically double-peaked,
at a height of 2.8 in units of the nearby continuum, with a weak central
depression, and a total equivalent width of about 13\,\AA\ in emission,
completely filling in the photospheric wings of
\mbox{H}$\mathrm{\alpha}$. Overall, the spectrum is very typical of an early
Be star seen at low to intermediate inclination.

A period analysis for the full 4\,yr Kepler data of \mbox{ALS}10705 and \mbox{StH$\alpha$}166 has been
described by \citet[][their Sect.~11]{2015MNRAS.451.1445B}. As the first part of the analysis
here, a similar technique is employed, the Lomb-Scargle analysis, and the
results are the same.

\subsection{\mbox{HD}186567 (\object{KIC11971405})}
\mbox{HD}186567 is considered as an example for a Be star by
\citet{2015MNRAS.450.3015K}. The original source of this classification is
unclear, but since that the star shows several Be-star typical outbursts
\citep[see Fig.~\ref{fig:trends} vs.][for a theoretical model how such
  outbursts produce the observed light curve]{2012ApJ...756..156H} there is no
strong reason to doubt the classification.

\mbox{HD}186567 is a late type B star for which \citet{2014ApJS..211....2H} give a
temperature of about 12\,kK and $\log g$ of 3.7.  \citet{2009AcA....59...33P}
again identify the star as aperiodic variable in ASAS data, while
\citet{2012AJ....143..101M} classify the variations as hybrid
SPB and $\mathrm{\beta}$~Cep showing frequency groups in the Kepler data, with a
  dominant value of about 4\,c/d.

\subsection{Light curve preparation}

For these three stars all available Kepler data were retrieved from MAST,
taken from May 2009 to May 2013 in 18 ``quarters'', a term which is used to
distinguish observing setups of the satellite. For details see Section 4 of
\citet{2011MNRAS.413.2403B} and the references therein. As date the
``Barycentric Kepler Julian Date'' is used, defined as $\mathrm{BKJD} =
\mathrm{BJD} - 2\,454\,833$, where $\mathrm{BJD}=2\,454\,833$ corresponds to
Jan. 1, 2009. All fluxes were converted to a magnitude scale. As data from
different quarters may not join smoothly, they were pre-processed for each
quarter individually by subtracting a linear trend before merging the quarters
into a single data set.

In addition to instrumental trends, and apart from short-term periodic/cyclic
variations, which are in the focus of this work, Be stars may show medium to
long-term secular variability due to radiative processes (see
\citeauthor{2012ApJ...756..156H}, op.\ cit.) in their circumstellar
environment. Their gaseous disks form and dissipate on time scales of
sometimes weeks, but more often months and even decades. The resulting
variability in the visual domain can surpass the amplitude of the short-term
variations by as much as three orders of magnitudes in extreme cases. In the
three stars under investigation, this ratio is considerably more benign, less
than one order of magnitude. For the most variable star in the sample,
\mbox{StH$\alpha$}166, the peak-to-peak amplitude of the secular variations is
0.1\,mag, and for the strongest periodic variation it is 0.016\,mag.

In terms of data processing it is irrelevant whether such a trend is (partly)
stellar or (partly) instrumental; it can be removed in a single step. As a
tool for time series analysis the {\sc vartools} package, published by
\citet{2008ApJ...675.1254H}\footnote{v1.32, available at\\ {\tt
    http://www.astro.princeton.edu/$\sim$jhartman/vartools.html}} is used. The
removal of secular trends was implemented as a Fourier-based high-pass filter.

\begin{figure*}[t]
\begin{center}
\includegraphics[angle=0,width=18cm,clip]{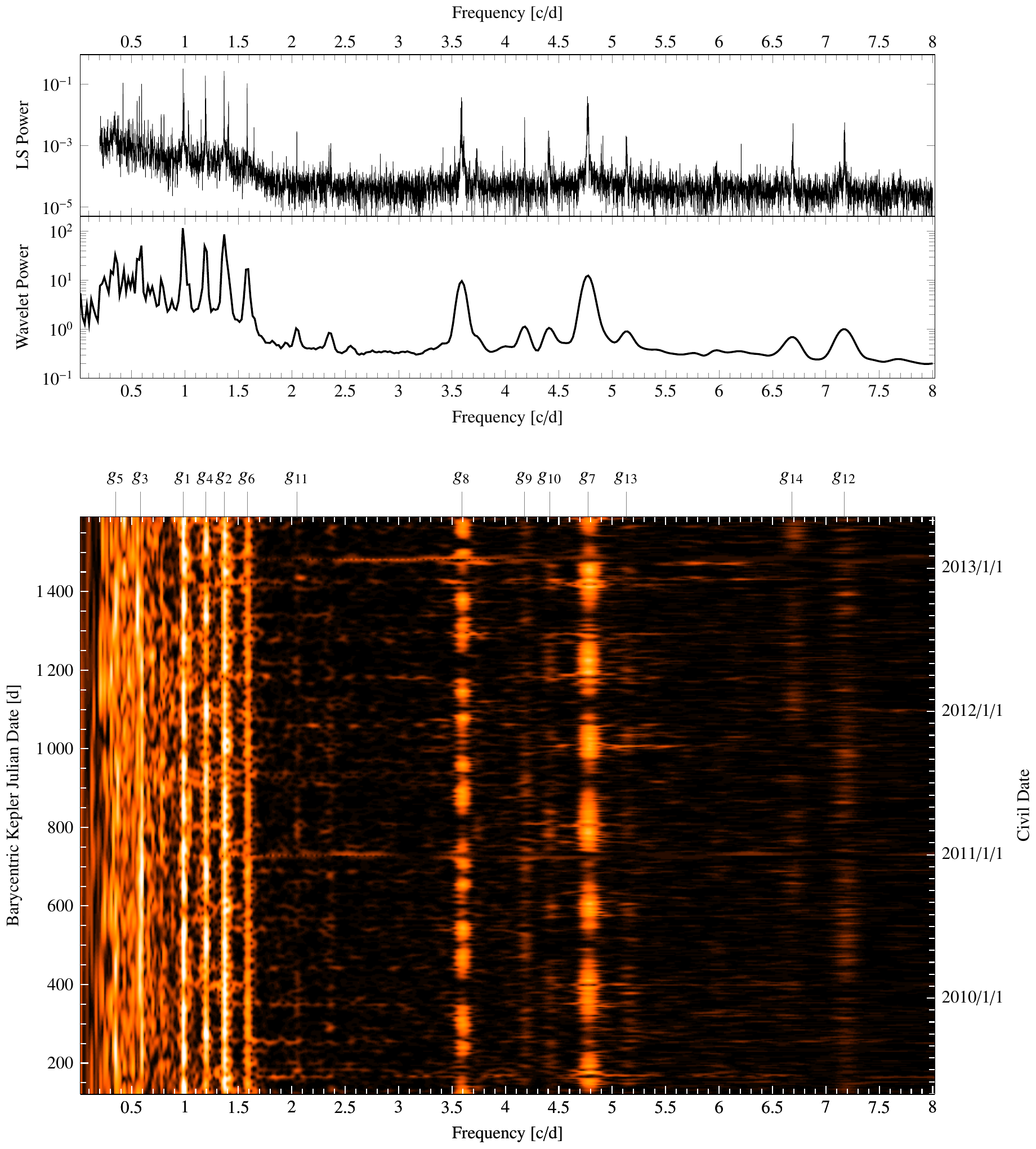}%

%
%
%
\end{center}
\caption[xx]{\label{fig:KIC008per} Time series analysis of
  \mbox{ALS}10705. From top to bottom the LS periodogram, the time averaged
  result of the wavelet analysis, both shown in logarithmic scale, and the
  full 2D result of the wavelet analysis, where the intensity is displayed in
  logarithmic scale, too. The frequency groups discussed in the text are
  identified above the lowermost panel.}
\end{figure*}

\begin{figure*}[t]
\begin{center}
\includegraphics[angle=0,width=18cm,clip]{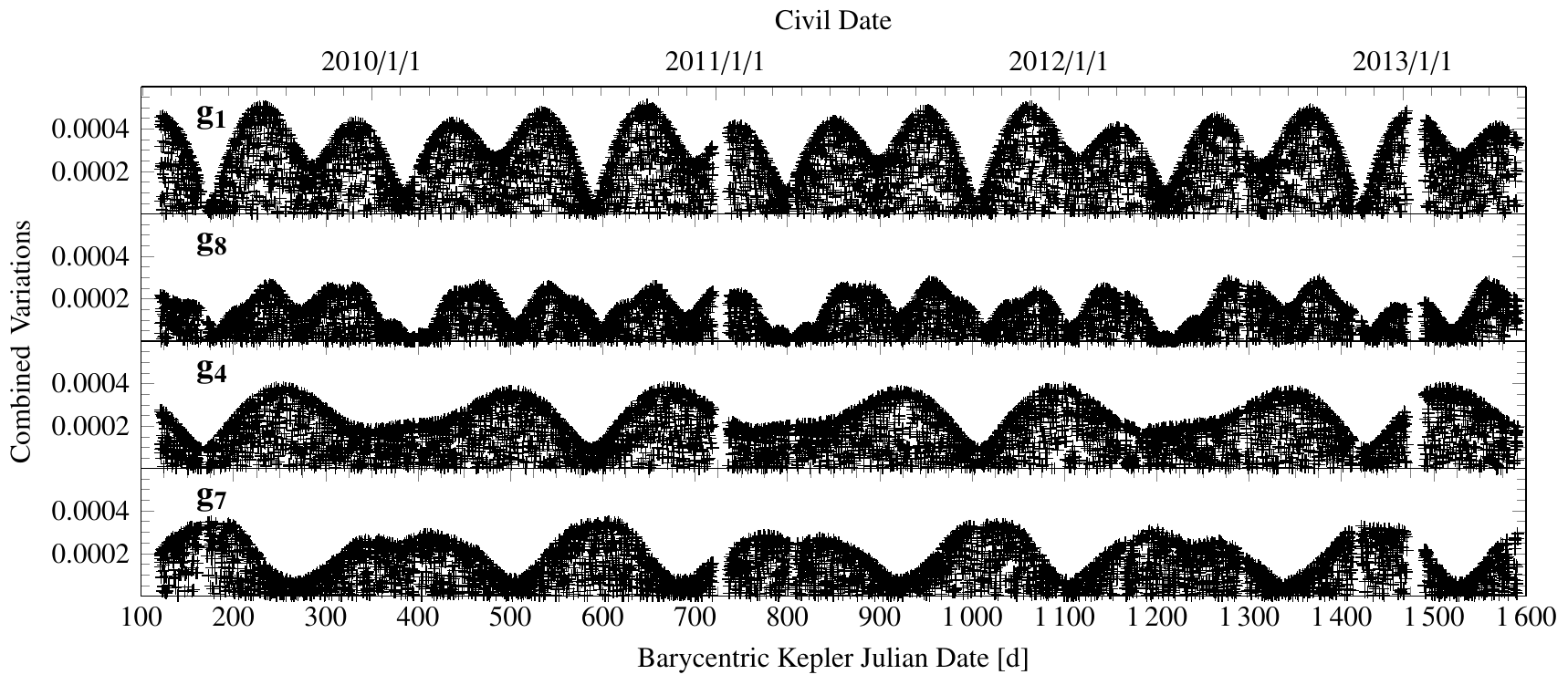}%

\end{center}
\caption[xx]{\label{fig:KIC008pow} Amplitude variation with time for $g_1$,
  $g_8$, $g_4$ and $g_7$ (top to bottom) of \mbox{ALS}10705. The individual
  points were computed by co-adding the fits to all frequencies found by LS
  belonging to each group, and then the absolute value was plotted. The upper
  envelope of each group, therefore, represents the amplitude variations seen
  in the lower panel of Fig.~\ref{fig:KIC008per} and illustrates the various
  correlations described in the text. }
\end{figure*}

The precise method was to find frequencies in the range between $10^{-3}$ and
0.2\,c/d with the generalized Lomb-Scargle (LS) periodogram
(\citealt{2009A&A...496..577Z}, implemented using
\citealt{1992nrca.book.....P}). These frequencies are removed iteratively by
fitting a sinusoid to the actual data (pre-whitening, explicitly without
removing harmonics). The iterative removal is conducted at least down to, and
mostly beyond the noise level, which is apparent as a frequency continuum in
the LS-periodogram. Typically hundreds of frequencies need to be removed
before reaching that stage.

A major factor in the applicability of such a method are the aliasing
properties of Kepler data, and in fact other space mission data, too: There
are no strong peaks of the window function in the range of typical Be star
frequencies, which are about 0.5\,c/d to about 10\,c/d. For this reason, any
removal of power at frequencies below 0.2\,c/d, that is applying a frequency
high-pass filter, does not affect the power spectrum at frequencies of
astrophysical interest. Mathematically speaking, therefore, such filtering is
not even required; results of the subsequent period search are the same,
whether the high-pass filter is applied or not. However, in order to obtain a
better visual representation of the secular vs.\ the periodic behavior (see
Fig.~\ref{fig:trends} for an example), and to fold the data with the obtained
periods, the high-pass filter is applied in any case.

{\begin{table}[b!]
\caption{\label{tab:KIC08}Comparison of frequency groups found in the wavelet
  vs.\ the dominant frequency of each group found in the LS analysis of
  \mbox{ALS}10705. For LS the frequency found first for each group, that is the
  strongest one, is given, together with the iteration number. The difference
  between the two values is not indicative for the precision, which is much
  higher.}
\begin{center}\small
\begin{tabular}{lrr|rrr}
&\multicolumn{2}{c}{Wavelet}&  \multicolumn{3}{c}{Lomb-Scargle} \\
Name &Freq.   & Power  & Freq.   & Semi-ampl. & Iter.\\
&{[c/d]} &        & [c/d]   & [mmag]     &  \#   \\
\hline
 $g_1$  &  0.984  &  115.1  &    0.981  &     0.31    &   2\\
 $g_2$  &  1.368  &   85.4  &    1.366  &     0.33    &   1\\
 $g_3$  &  0.585  &   50.6  &    0.594  &     0.18    &   6\\
 $g_4$  &  1.192  &   49.3  &    1.196  &     0.10    &  12\\
 $g_5$  &  0.351  &   32.6  &    0.344  &     0.07    &  21\\
 $g_6$  &  1.582  &   16.8  &    1.582  &     0.18    &   5\\
 $g_7$  &  4.772  &   12.3  &    4.767  &     0.10    &  11\\
 $g_8$  &  3.590  &    9.6  &    3.589  &     0.12    &   8\\
 $g_9$  &  4.178  &    1.1  &    4.180  &     0.05    &  37\\
 $g_{10}$  &  4.415  &    1.1  &    4.402  &     0.03    & 107\\
 $g_{11}$  &  2.048  &    1.0  &    2.047  &     0.04    &  53\\
 $g_{12}$  &  7.172  &    1.0  &    7.176  &     0.04    &  61\\
 $g_{13}$  &  5.132  &    0.9  &    5.135  &     0.03    & 124\\
 $g_{14}$  &  6.681  &    0.7  &    6.693  &     0.04    &  69\\
\end{tabular}
\end{center}

\end{table}}

Fig.~\ref{fig:trends} shows the original data after removing mean value and
linear trends, together with the curve obtained by the high-pass filter (i.e.,
the sum of all low-frequency terms iteratively found). The variability of
\mbox{StH$\alpha$}166 has strong secular outbursts, while the variations of
\mbox{HD}186567 are dominated by short term variations, which remain in this
stage, with only few outbursts towards the end of the observing time span. The
variations of \mbox{ALS}10705 are not obviously identified as either stellar
or circumstellar. Recalling the fact that the star was without emission, and
considering the rather low amplitude and long time scale of the secular
variability, they are likely instrumental effects, also since the pattern
roughly repeats annually. If the other stars suffer from a similar pattern, it
is hard to see because their variations are dominated by stellar secular
variability. In any case, these long-term trends are subtracted in this step,
and therefore of no consequence for the results below.

\subsection{Light curve analysis}

The final analysis is done in two different ways. First, the data are analyzed
using the generalized LS periodogram, just as described above, but now in the
frequency range of astrophysical interest from 0.2\,c/d to 25\,c/d; the latter
is about the Nyquist frequency of the long cadence data (24.469\,c/d). As this
method is the traditional way of analyzing such a light-curve, it yields
essentially identical results to the ones described by
\citet{2011MNRAS.413.2403B} and \citet{2015MNRAS.450.3015K}.

As an alternative way of looking at the data a wavelet analysis is used, again
in the implementation included in the {\sc vartools} package by
\citet{2008ApJ...675.1254H}, based on \citet[][see there also for
  examples and a critical assessment of the method]{1996AJ....112.1709F}. In
the most simple terms, a wavelet is an oscillating function, here a sinusoid,
with a decay term:
\begin{equation}
\exp(2 i \pi f (t-t_0) - c (2 \pi f)^2 (t-t_0)^2),
\end{equation}
where $f$ is the frequency and $c$ a scaling factor of the decay.  The decay
term is kept constant in units of the frequency, that is the wavelet at any
given frequency sees only a fixed number of cycles towards the past and
the future; the effective time base of the analysis decreases as frequency
increases. As a consequence, the frequency resolution of a wavelet analysis is
not a function of the total time base, but one of the frequency sampled: The
higher the frequency, the lower the resolution. In this work an unusually slow
decay was used. While a standard value is $c=1/8\pi^2\sim0.0125$, here
$c=0.0001$ was adopted, i.e., coherence over a few hundred instead of just a
few oscillations. This was done first since such a rapid change of amplitude
and period is not expected, and second to increase the frequency resolution to
a level where most of the various groups can be safely disentangled.

For visualization, a 2D representation is used for the result. For
any given time step (here: one day) a wavelet periodogram is obtained. All
periodograms are then stacked into an image with frequency on the x-axis, time
step of the wavelet analysis on the y-axis, and power on the z-axis (the
latter coded as intensity in the figures here).

\section{Results}  \label{sec:results}

\subsection{\mbox{ALS}10705}
In terms of variability, \mbox{ALS}10705 is the least complicated object in
the sample. Since it is an inactive Be star without any spectroscopic sign of
a disk, and even the very low amplitude secular photometric variations are
likely to be instrumental, one can be quite certain that there is no
circumstellar variability or disk feeding complicating the picture, and
therefore all variability detected is photospheric, making \mbox{ALS}10705 a
valuable case to compare variations in other, more active, Be stars to.

First concentrating on the LS periodogram in Fig.~\ref{fig:KIC008per}, one
would classify \mbox{ALS}10705 as a hybrid pulsator, since both low and high
frequencies are present, often taken as indicative of $g$- and $p$-modes,
respectively.  We ignore for the moment that such a simple classification is
questionable in the case of rapidly rotating objects such as Be stars, where
rotation can alter the observed frequencies considerably, and defer this issue
to the discussion in Sect.~\ref{sec:rot}.

Already in the LS periodogram it can be seen immediately that the variation at
about 3.6 and 4.8\,c/d cannot be described by a single, stable frequency, but
there are a number of peaks, separated in distinct groups. These would
normally be interpreted as beating frequencies, that is a number of nearby phase
coherent and amplitude stable frequencies that, when co-added to each other,
evince the impression of a frequency with variable amplitude, becoming
stronger and weaker with a function called the beating envelope. When two
individual beating frequencies remain coherent during the sampled time base
and are of equal amplitude, the envelope varies strongly between zero and
twice the amplitude of an individual frequency.  A closer inspection of the
frequencies below 2\,c/d reveals that the same is the case here, each single
peak dissolves to a number of sub-peaks, forming a group of frequencies with a
strong beating pattern. Such patterns are seen in Fig.~\ref{fig:KIC008pow},
where the absolute values of the co-added { sinusoidal fits to the
  individual frequencies} are shown for several groups, { so that the total
  amplitude and its variations are represented by the upper envelope of each
  group.}

At frequencies lower than 2\,c/d increasing noise is seen.  When
pre-whitening the LS-periodogram, one finds that even after more than 100
iterations significant power still remains in the power spectrum, with clear
peaks belonging to one of the above described groups. The identified groups
are listed in Table~\ref{tab:KIC08}.

In the time average of the wavelet transform, these individual groups are not
resolved. The decrease of frequency resolution with increasing frequency is
obvious, and the residual of the periodic power that has been removed by the
high-pass filtering is clearly apparent at frequencies below 0.2\,c/d.  Except
for being of lower quality, however, this plot does not really offer new
insight beyond the LS periodogram. Rather than the myriad of peaks seen in the
LS periodogram, one can identify individual peaks in the averaged wavelet
transform. These correspond to the groups mentioned above (see
Fig.~\ref{fig:KIC008per}).

In the full 2D representation of the wavelet analysis, other, and partly
different aspects are more easily apparent to the eye. Among other things, the
effect of data outliers is easily seen as a sudden, short lived surge of power
at all frequencies, akin to the Fourier transform of a $\delta$-function,
which displays constant power at all frequencies (see
e.g., Fig.~\ref{fig:KIC008per}, lower panel at $t=725$). Another difference is
scientifically far more interesting. Instead of resolving a frequency group
into a high number of single frequencies, that reproduce the over-all behavior
of the group by their beating, the wavelet transform sees only a single
frequency with variable power. Or, in other words, at the expense of frequency
resolution it visualizes the phase and strength of the envelope function of
the -- supposed -- beating, something completely lost in the traditional
visualization of a 1D periodogram.  This opens the view for new and
alternative interpretations.

\begin{figure*}[t]
\begin{center}
\includegraphics[angle=0,width=18cm,clip]{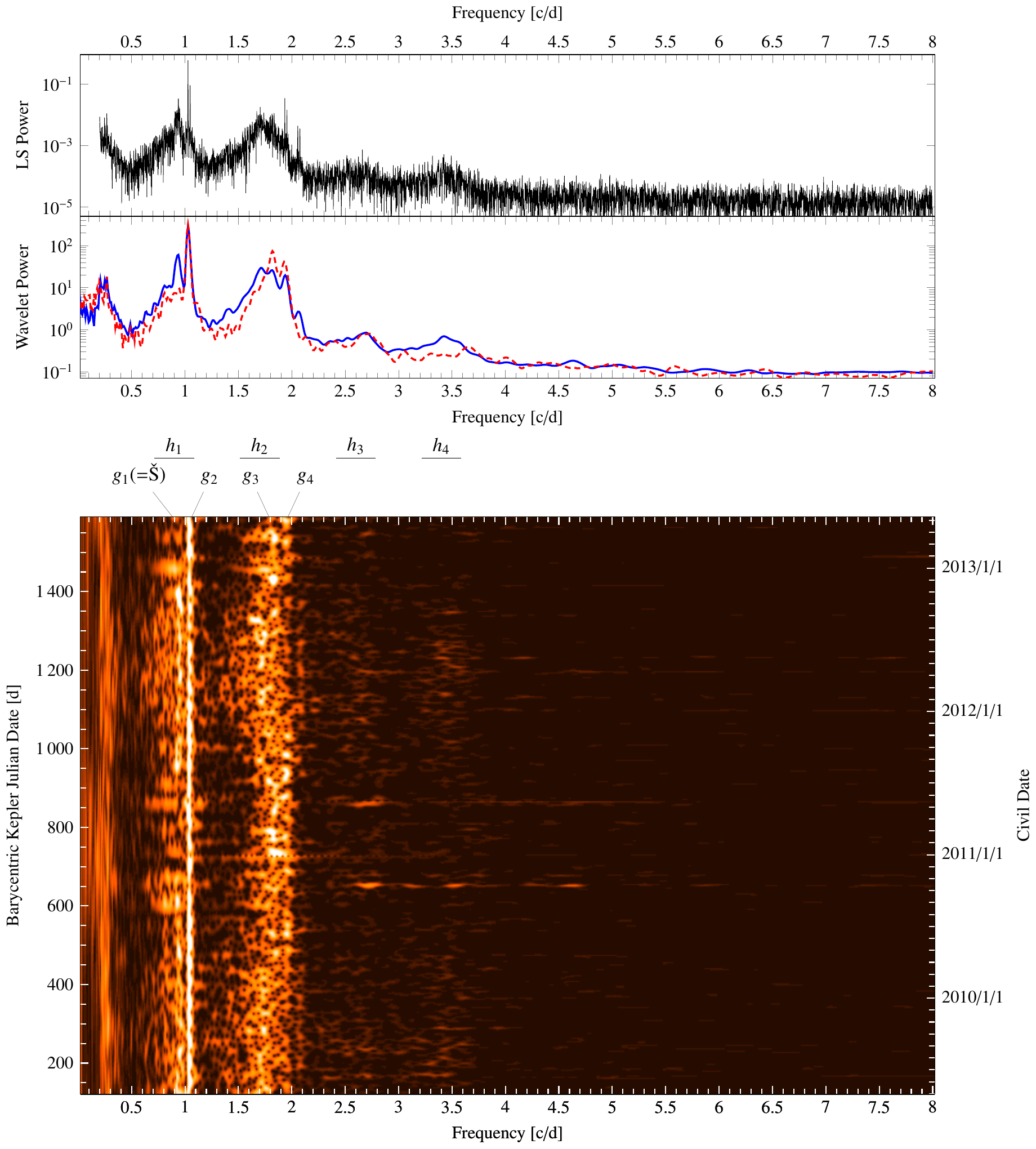}%

%
%
%
\end{center}
\caption[xx]{\label{fig:KIC006per} Like Fig.~\ref{fig:KIC008per}, but for
  \mbox{StH$\alpha$}166. In the middle panel, instead of the overall wavelet
  power spectrum, the average power spectrum during more variable phases
  ($t<700$ and $900<t<1500$) is shown in as a solid line, during less variable
  ($700<t<900$ and $t>1500$) as a dashed one. The label \v{S} denotes the
  circumstellar \v{S}tefl frequencies (see Sect.~\ref{sec:stefl}).}
\end{figure*}

Addressing the envelope of the amplitude for each frequency group $g_i$ as
$A_i$,  Fig.~\ref{fig:KIC008pow} helps to visualize $A_1$, $A_8$, $A_4$, and
$A_7$ from top to bottom. The following observations can be made about these
envelopes:
\begin{itemize}
\item $A_1 \sim A_8$ and $A_4 \sim A_7^{-1}$.
\item $A_1$ has minima when either $A_4$ or $A_7$ have a
  minimum. $A_1$ minima coinciding with $A_4$ minima are deeper than those
  coinciding with $A_7$ minima. 
\item In Fig.~\ref{fig:KIC008per} it can as well be seen that $A_7 \sim A_{10}
  \sim A_{13}$. Curiously, also { the frequency difference $g_7-g_{13}$
    equals that of} $g_{10} - g_7$. A relation between $A_4 \sim A_9$ is there
  as well, though it is probably more obvious to the eye when described as
  $A_9\sim A_7^{-1}$.
\item There is no obvious pattern relation for $A_2$, $A_3$, $A_5$, or
  $A_{6}$, however the amplitudes are variable.
\item  For $A_{11}$, $A_{12}$ and $A_{14}$  no pattern can be identified, as
  they are barely above the noise.

\end{itemize}
Summarizing the results for \mbox{ALS}10705, one can describe the photometric
variability in two different frameworks: In the Lomb-Scargle Fourier
framework, the star has fourteen frequency groups, each with a dozen and more
frequencies above the noise, that produce a complicated beating
pattern. Alternatively, in the wavelet framework, the star has fourteen single
frequencies, that have strongly time variable amplitudes, and in which the
amplitude variations are correlated between some of the frequencies.

\subsection{\mbox{StH$\alpha$}166}\label{sec:sth}

This object is, in a sense, the opposite of \mbox{ALS}10705: it is an active
Be star with perpetual outbursts, that is ongoing disk feeding. This does not
mean the outbursts are equally strong at all times, however. In
Fig.~\ref{fig:trends} it is seen that the outbursts never fully cease, but
from $t=700$ to 900, and again after $t=1500$ they are less strong and less
frequent than at other times. 

Even after removing the long-term variations, mostly due to the feeding
outbursts, a greater peak-to-peak variability of the residuals remains at
times when outburst have been stronger. In the middle panel of
Fig.~\ref{fig:KIC006per} the average wavelet power over the more (less)
feeding phases is plotted as a solid (dashed) line.
 
In order of increasing frequency value, the groups are designated as $g_1$ at
0.94\,c/d, $g_2$ at 1.03\,c/d, $g_3$ at 1.82\,c/d, and $g_4$ at 1.93\,c/d (see
Fig.~\ref{fig:KIC006per}). Further there are four very broad bumps of
power, which are centered at roughly 0.9, 1.7, 2.6, and 3.4\,c/d ($h_1$ to
$h_4$, respectively).

\begin{figure}[t]
\begin{center}
\includegraphics[angle=0,width=8.8cm,clip]{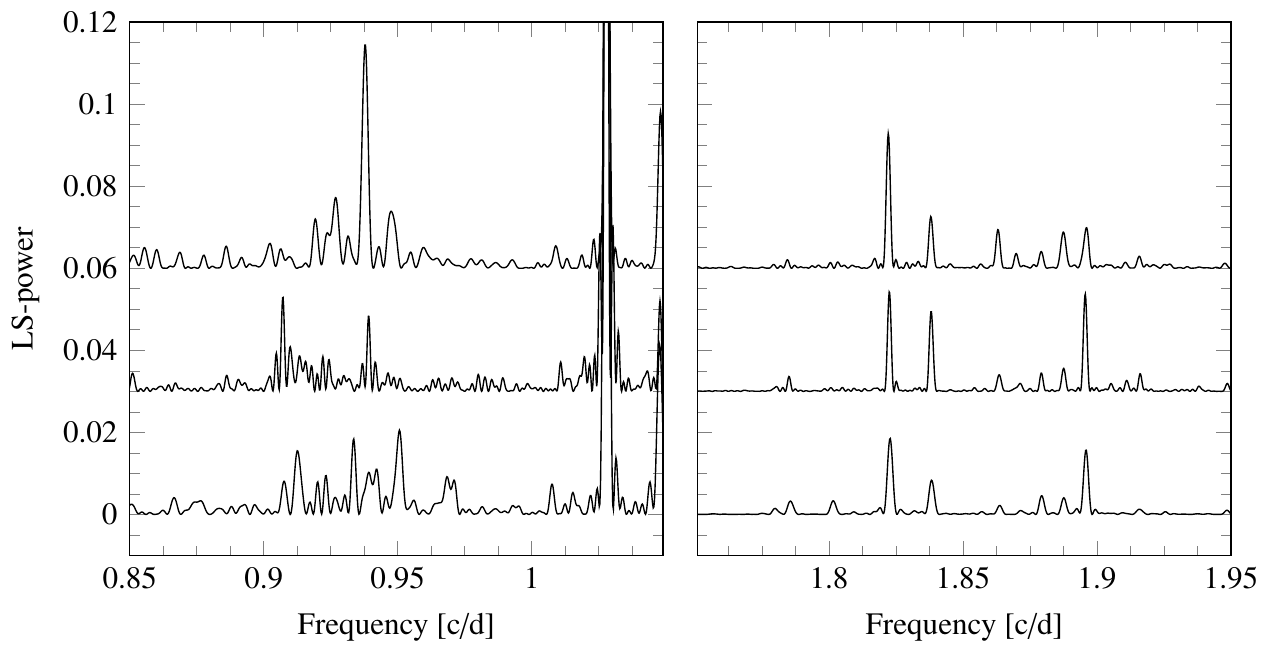}%
\end{center}
\caption[xx]{\label{fig:quarters} LS power spectra of the data divided into
  three sub-runs; from top to bottom the independent analysis for quarters
  12--17, 6--11, and 0--5 is shown. Left: \mbox{StH$\alpha$}166 in the region
  of the \v{S}tefl frequencies $g_1$ around 0.95\,\c/d; some part of $g_2$ is
  seen above 1\,c/d. Right: The same for for \mbox{HD}186567 in the region of
  $g_2$.}
\end{figure}

Of the more sharply defined frequency groups, $g_1$ is only apparent in the
more variable phase and appears to be what is called the \v{S}tefl frequency
of this star (see Sect.\ref{sec:stefl}). In the LS analysis it shows as a
rather broad distribution of numerous single frequencies (see
Fig.~\ref{fig:quarters}). In contrast, $g_2$ is persistently present and much
sharper defined in the LS periodogram. The frequency of $g_1$ is about 10\%
lower than that of $g_2$. We note here that this combination is very
distinctive for Be stars in outburst, but will address it further only in the
discussion in Sect.~\ref{sec:stefl}.

In order to see whether there is at least some preferred, persistent frequency
in this group $g_1$, the data was divided in three substrings that were
analyzed individually. The individual LS power spectra are shown in the left
panel of Fig.~\ref{fig:quarters}. The frequencies present in the individual
sub-strings are entirely different from each other, that is there is no
long-term coherency or frequency stability in this group. The power
spectrum shown in the middle is computed from data taken mostly in the time of
lower disk feeding ($ 539.5 < t < 1098.3$), and shows much less total power
than the two other ones.

$g_3$ and $g_4$ are situated in the wing of the strongest broad bump. Due to
their weakness they are actually dominated by the bump. Close inspection of
the 2D periodogram, and the middle panel of Fig.~\ref{fig:KIC006per}, suggests
at least two general statements, namely first that $g_3$ and $g_4$ are
correlated, and second that they seem to be stronger in the less variable
phases {i.e., anti-correlated with the strength of $g_2$}. A fifth group
  was not caught by the LS-analysis, because its total amplitude was less than
  that of the bumps, but is apparent to the eye at about 2.1\,c/d. This group
  is stronger in the more variable phases.

The most obvious features in the power spectrum are not the discrete frequencies
or groups, however. Rather, it is dominated by broad bumps of power. The
two strongest of these bumps were already identified by
\citet{2011MNRAS.413.2403B} as the dominant feature of the
spectrum. \citeauthor{2011MNRAS.413.2403B} mention a third peak at 0.1\,c/d,
but in the interpretation here that peak is due to the low frequencies
representing the long-term outburst and their harmonics that were removed in
the pre-processing steps (see Fig.~\ref{fig:trends}). It has nothing to do
with the bumps at higher frequencies.

In addition to the bumps $h_1$ { at 0.9\,c/d} and $h_2$ already mentioned
by \citet{2011MNRAS.413.2403B}, both the LS and the wavelet analysis show two
further bumps at higher frequencies { $h_3$ and $h_4$}, more prominent when
the disk feeding activity is high. As the wavelet analysis makes clear, these
are not  groups of real frequencies.  Even in the LS analysis,
pre-whitening for more than one hundred frequencies within these bumps does
not alter their appearance or power in any significant way. This clearly
indicates aperiodic, at best short-term cyclic variations on a scale of about
0.9\,c/d. The higher bumps $h_2$ to $h_4$ are at positions that make them
likely to be a sign of the non-sinusodiality of this aperiodic variability. In
particular, $h_1 \approx h_2/2 \approx h_3/3 \approx h_4/4$. If $h_1$ was due
to true frequencies, $h_2$ to $h_4$ would be their harmonics. However, as they
are not coherent frequencies, instead of being real harmonics, one may
consider the bumps at higher frequencies as ``group'' or
``statistical'' harmonics (see also Paper I). Another distinctive feature of
$h_1$ and $h_2$ is that they become broader towards lower frequencies at the
time of outbursts.

\begin{figure*}[t]
\begin{center}
\includegraphics[angle=0,width=18cm,clip]{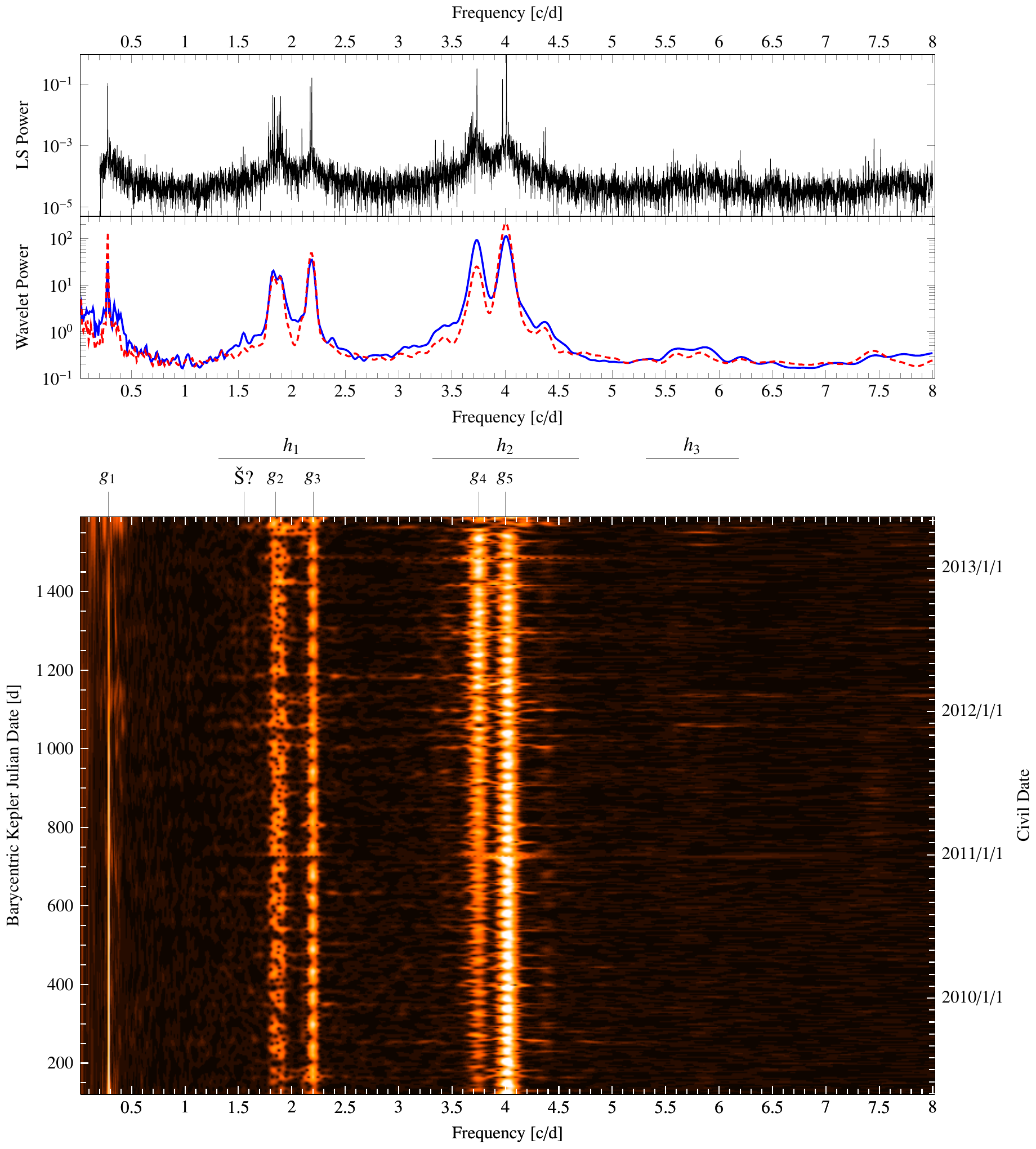}%

%
%
%
\end{center}
\caption[xx]{\label{fig:KIC011per} Like Fig.~\ref{fig:KIC006per}, but for
  \mbox{HD}186567. All data before $t=1000$ are averaged into the less
  variable wavelet power spectrum (dashed), data after into the more variable
  one (solid). The label \v{S} denotes the possible circumstellar \v{S}tefl
  frequency (see Sect.~\ref{sec:stefl}), seen better in the average power
  spectra rather than in the 2D representation.}
\end{figure*}

\begin{figure*}[t]
\begin{center}
\includegraphics[angle=0,width=18cm,clip]{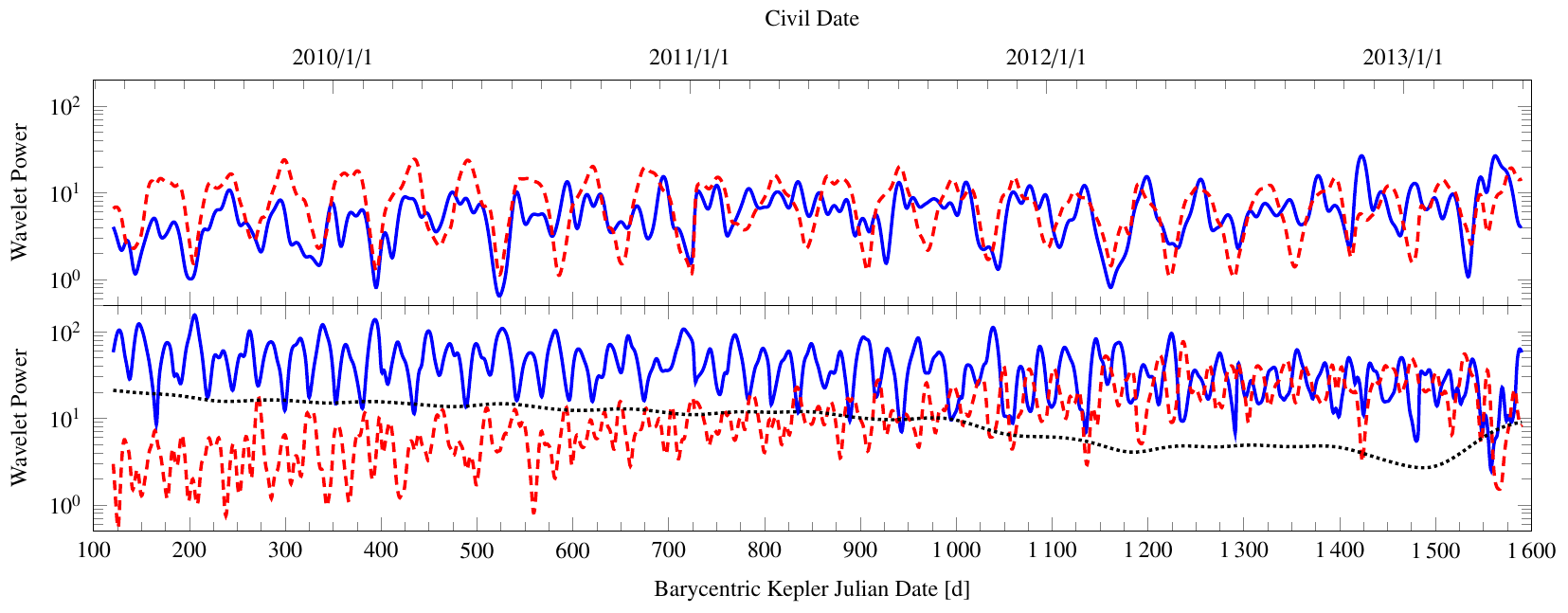}%

\end{center}
\caption[xx]{\label{fig:KIC011pow} Amplitude variation with time for \mbox{HD}186567. 
Curves were created by averaging the lowermost panel of Fig.~\ref{fig:KIC011per}
over the respective frequency regions.
Top panel    shows  $A_2$ solid and  $A_3$ as a dashed line.
Bottom panel shows  $A_1$ dotted  $A_4$ dashed, and  $A_5$ as a solid line.
}
\end{figure*}

\subsection{\mbox{HD}186567}
In terms of both Be star activity and disk feeding, this star is an
intermediate example between \mbox{ALS}10705 and \mbox{StH$\alpha$}166. There
are no detectable outbursts in the first two years of the observations, but
the star shows increasing outburst variability after $t=1000$. Due to the lack
of spectroscopic data one cannot tell whether the star was completely without
emission before $t=1000$, or rather in a disk dissipation phase, in which the
circumstellar disk is still present, but no longer being fed by outbursts.

The power spectrum in both LS and wavelet analysis of \mbox{HD}186567 (see
Fig.~\ref{fig:KIC011per}) shows four clear frequency groups, and a number of
isolated peaks. Again several correlations between the amplitude envelopes of
each group exist.

The strongest features are the groups at $g_2=1.8$\,c/d, $g_3=2.2$\,c/d,
$g_4=3.7$\,c/d, and $g_5=4$\,c/d, plus the peak at $g_1=0.28$\,c/d, numbered
in the order of increasing frequency. In fact, also $g_1$ is resolved by the
LS analysis into four closely spaced frequencies, which is, however,
negligible compared to the dozens of frequencies found in the other groups.
{ The frequency value of $g_1$ is about the difference between $g_4$
  and $g_5$.}  Although $g_4\approx2\times g_2$, it is different from the
above mentioned statistical harmonics in the sense that here the behavior
of $A_4$ is not correlated with $A_2$.

What is immediately seen is that the amplitude envelopes $A_4$ and $A_5$ are
anti-correlated on the longest time scale observed, that is years. While $A_4$
gains in power from 2009 to 2013, $A_5$ decreases towards the end. Not obvious
from Fig.~\ref{fig:KIC011per}, but well seen in Fig.~\ref{fig:KIC011pow} is
that $A_1$ follows the same trend as $A_5$, slowly decreasing in power. $A_2$
vs.\ $A_3$ shows a similar long-term behavior, but much less pronounced: While
$A_2$ increases somewhat, $A_3$ decreases a little (see as well
Fig.~\ref{fig:KIC011pow} rather than Fig.~\ref{fig:KIC011per}).

On short times scales, there are as well correlations between the amplitude
envelope curves. $A_2$ and $A_3$ are quite well positively correlated;
certainly not perfect but clearly beyond what one would expect from random
variations. $A_4$ and $A_5$, however, are anti-correlated, that is $A_4$ being
strong means $A_5$ is weak and vice versa. These correlations are seen because
each group has a short term variability in itself, in principle like the ones
seen in ALS10705, but much faster (see Fig.~\ref{fig:KIC008pow} for
comparison).
 
In the overall power spectra, it is also seen that there are two very broad
bumps, at about $h_1=2$\,c/d and at about $h_2=4$\,c/d. A third one, $h_3$
might be identified at about 5.5 to 6\,c/d, stronger when there is disk
feeding.  $g_2$ and $g_3$ seem to be symmetrical around the center of $h_1$,
and $g_5$ centered on $h_2$. The behavior of the bumps is very similar to the
ones in \mbox{StH$\alpha$}166. Before $t=1000$, when no signature of disk
feeding is seen, they are weaker, and also a bit shifted towards higher
frequencies, while when the disk is being fed after $t=1000$ they are stronger
and slightly shifted towards lower frequencies.

In detail, the variability properties of $g_2$ look somewhat different from all
that was described so far. Although $A_2$ is correlated with $A_3$, even in
the wavelet analysis it is not possible to identify a single, stable frequency
that would just have variable amplitude.  \citet{2015MNRAS.450.3015K} state
that the period values in a periodogram computed over observing quarters 0 to
9 are the same as the ones over quarters 10 to 17. We have repeated this
exercise with quarters 0 to 5, 6 to 11, and 12 to 17, that is with a somewhat
finer sampling, less susceptible to averaging. The same approach was already
applied to \mbox{StH$\alpha$}166 (see Fig.~\ref{fig:quarters}). The periods
are indeed quite stable, although their formal differences exceed 3$\sigma$
somewhat.  The strongest of those frequencies, at 1.822\,c/d, has an average
semi-amplitude of about 1\,mmag.  Inspecting Fig.~\ref{fig:KIC011per} in the
region of $g_2$ more closely, one finds the power distributed over a frequency
range that is broader than the resolution of the wavelet analysis at this
frequency, but still much narrower than the broad bumps.  The adjacent pattern
of $g_3$ serves as a comparison how a single frequency with variable amplitude
(or two stable beating frequencies) would look like.

\section{Discussion}\label{sec:discussion}
\subsection{Amplitude correlations}

The commonly adopted interpretation of the LS periodogram of a B type star is
that of a large number of independent frequencies, which, if coming in groups,
produce beating pattern with very low frequencies that show as slowly varying
amplitude envelopes. This work offers nothing to actually disprove this
interpretation. However, the observed correlations between the amplitude
envelopes of supposedly independent frequency groups are not easily explained
in such an interpretation; other explanations, in which such correlations
arise more naturally, should be explored. Most typical is a negative
correlation between two strong frequency groups (\mbox{ALS}10705: $A_7/A_4$,
\mbox{HD}186567: $A_4/A_5$, possibly \mbox{StH$\alpha$}166: $A_3/A_4$), and a
strong frequency group might be accompanied by a number of weaker ones with
positive correlation (\mbox{ALS}10705: $A_7+A_{10}+A_{13}$, $A_4+A_9$,
$A_1+A_8$). However, more complicated relations are observed as well
(\mbox{ALS}10705: $A_1$ vs.\ $A_7/A_4$).  Such relations prompt the question
whether there are internal processes in the star that redistribute energy
between pulsation modes on short time scales back and forth. Resonant mode
coupling has been described much earlier, e.g., by
\citet{1982AcA....32..147D}, but so far the idea was applied mostly to
$\delta$\,Sct and RR\,Lyr stars, rarely to B stars.

\subsection{The \v{S}tefl frequencies of \mbox{StH$\alpha$}166}\label{sec:stefl}

In Sect.~\ref{sec:sth} $g_1$ at 0.94\,c/d of \mbox{StH$\alpha$}166 is
described as transient and unstable in frequency and phase. This and the
presence of a more stable variation $g_2$ at a slightly higher frequency of
1.03\,c/d hallmarks it as a phenomenon often found to be associated with disk
feeding, and $g_2$ as the main stellar pulsational frequency that would be
detectable by spectroscopy. Such frequencies have first been described by
\citet{1998ASPC..135..348S}. Then they were named ``transient frequencies'' in
reference to their non-persistent nature. For a detailed description of
\v{S}tefl frequencies see Sect.~2 of Paper I.

From Fig.~\ref{fig:quarters} it is clear that there is no single such
frequency; rather these are fairly short-lived individual episodes of cyclic
variability, which occur within a narrow frequency region of about
$\pm5\%$. Paper I, based not only on the two stars studied there, but as well
re-assessing the literature, has made a strong case that this variability is
directly related to the feeding of fresh stellar material into the base of the
disk. Certainly the disk feeding pattern, strong in quarters 0--5 and 12--17,
but weak in 6--11, agrees with this. Some part of the variability identified
above as groups $g_3$ and $g_4$ could actually have some 
{ harmonic relation to $g_1$. As will be shown in forthcoming papers of
  this series, it often is the case in Be stars that the first harmonic of a
  circumstellar \v{S}tefl frequency is stronger than the base frequency
  outside of outbursts. The alternative would be that $g_3$ and $g_4$ are
  $g$-modes in their own right. In the same test as shown in
  Fig.~\ref{fig:quarters} $g_4$, just at different frequency values, is found
  to be stable in frequency, while $g_3$ is not. This means that while $g_3$
  might indeed be a harmonic of $g_1$, $g_4$ is more likely a real $g$-mode
  pulsation.}

One could argue from Fig.~\ref{fig:KIC011per} that also in the power spectrum
of \mbox{HD}186567 a \v{S}tefl frequency might be present after the star began
to feed the disk at about $t\approx 1000$, at a frequency value of about
1.55\,c/d. However, given the weak variability and the weak signature in the
power spectrum, that cannot be said with certainty. We note that there is nothing
like a \v{S}tefl frequency seen in the quiescent Be star ALS 10705.

\subsection{Broad bumps in the power spectra}

The broad bumps found in the power spectra of \mbox{StH$\alpha$}166 and
\mbox{HD}186567 are well known from other Be stars. They are typically
interpreted in terms of a dense population of low-order $g$-modes
\citep{2005ApJ...635L..77W,2007CoAst.150..213D,2008ApJ...685..489C}, though
\citet{2009AIPC.1170..339B} has suggested a rotational or circumstellar
interpretation. Here it is argued that they are indeed circumstellar, though
not necessarily connected to rotation, just due to aperiodic variability
residing in the circumstellar environment, that is the gaseous disk (see Paper
I). The bump at very low frequencies of up to about 0.1\,c/d is due to the
photometric variability of the long-term outbursts of Be stars. The bump at
the next higher frequency is suggested to be due to aperiodic variations on
about the time scale of that frequency, while bumps at still higher frequency
are (pseudo-)harmonics belonging to these aperiodic variations.

Variations on a scale of around 1 to 2\,c/d are ubiquitous in Be stars. It can
be $g$-mode pulsation, { stellar} rotation, or the orbital period in the
inner parts of the disk. However, neither a well defined frequency nor
coherence is found, and the position and strength of the bumps correlate with
the strength of the disk feeding. The latter is seen in the respective middle
panels of Figs.~\ref{fig:KIC006per} and \ref{fig:KIC011per}, where power
spectra are shown, averaged for more and less strong feeding phases. The
presence of a broad bump at all times in the wavelet analysis for
\mbox{StH$\alpha$}166 suggests that there are a large number of such
variations present at any given time there is a disk, while the diskless
\mbox{ALS}10705 is free of such bumps.  All these observations speak for a
purely circumstellar origin, rather than a stellar, photospheric one.

Unfortunately, the Be activity status of \mbox{HD}186567 during the Kepler
observations is not fully known. Clear  outburst were observed only
after $t\approx1000$, but also before the star may well have had a
circumstellar disk, either dissipating or sustained by minor outbursts
only. This would have been a disk in its dissipation phase, formed during
earlier outburst variability. Extended dissipation phases with no or very
little outburst variability are common in Be stars
\citep{2003A&A...402..253S,2014ApJ...786..120D}. In any case, in
\mbox{HD}186567 there was a growing number of outbursts after $t=1000$, and
the strength of the bumps $h_1$ and $h_2$ increased, while the previously
barely detectable $h_3$ became much clearer as outburst variability began.

\begin{figure*}[t]
\begin{center}
\includegraphics[angle=0,width=18cm,clip]{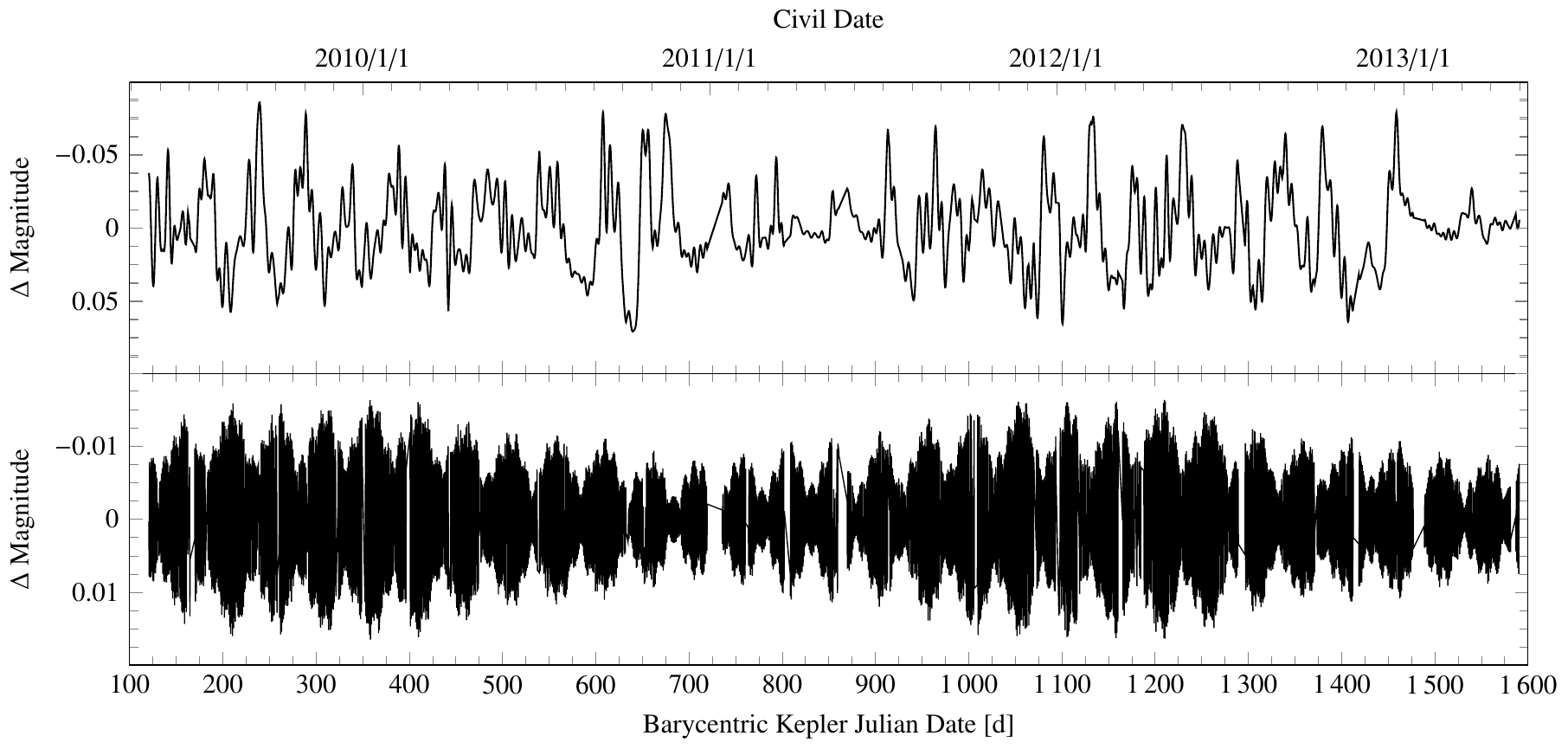}%
\end{center}
\caption[xx]{\label{fig:KIC06_f2} Combined variation of the $g_2$ group of
  \mbox{StH$\alpha$}166, reconstructed with the frequencies in Table~\ref{tab:KIC06_f2}
  (bottom) compared to the long-term outburst variability of the star (upper
  panel, see also middle panel of Fig.~\ref{fig:trends}).  }
\end{figure*}

\subsection{Outbursts and pulsation of \mbox{StH$\alpha$}166}
A detailed analysis of \mbox{StH$\alpha$}166 concerning the relation between
the outbursts and the pulsational behavior is out of the scope of this
work. However, already in the analysis here a relation akin to
\mbox{$\mathrm{\mu}$}~Cen \citep[see][]{1998cvsw.conf..207R} between the
combined amplitude of $g_2$ and the outbursts is seen (see
Fig.~\ref{fig:KIC06_f2}), i.e., amplitude maxima of $g_2$ co-incide with
outbursts. In the first 120 iterations of the LS analysis a total of nine
frequencies were found that belong to $g_2$
(Table~\ref{tab:KIC06_f2}). Independent of the interpretation as real
frequencies or just a mathematical representation of the overall evolution of
a single, true frequency, the co-added variations of those nine frequencies
describe the behavior of $A_2$, just as the co-added low frequencies describe
the secular behavior. The amplitude reconstruction is compared to the
long-term outburst evolution in Fig.~\ref{fig:KIC06_f2}. Both the long-term
evolution of the reconstructed amplitude, maxima at $t\approx350$ and
$t\approx1150$, as well as the short-term structure with intermediate maxima
every $\approx50$ days, and further sub-maxima even more often, agree well
with the temporal evolution of the outbursts of \mbox{StH$\alpha$}166. The
nine periods of $g_2$ (Table~\ref{tab:KIC06_f2}) can be grouped again in three
sub-groups seperated by about 0.02\,c/d, which is also the frequency value
where the by far highest power is found in the LS-analysis of the long-term
tends ($\approx0.02$\,c/d, or 50 days as mentioned above). This may be the
same type of behavior as seen in $\eta$\,Cen (see Paper I), where the
frequency difference of two close stellar frequencies is seen with high
amplitude not as a beating phenomenon, but as a frequency in its own right
with much higher amplitude than the two original frequencies.

\subsection{The $g_1$ and $g_2$ groups of \mbox{HD}186567}  


{ The frequency value of $g_1=0.28$\,c/d of this not very variable Be star
  of late B subtype is about the difference of two much higher frequencies,
  $g_4$ and $g_5$. A rotation period of about 3.5\,d would be unbelievably
  long for a Be star, which are thought to be rotating close to the critical
  limit. If $g_1$ is not a true pulsation mode, in which case it would have to
  be a retrograde $g$-mode, see Sect.~\ref{sec:rot}, it is a combination
  frequency of $g_4$ and $g_5$. Such frequencies have been found in Be stars
  (e.g., in $\eta$\,Cen, PaperI, or in $\omega$\,CMa, Sect.\ 4.1 of
  \citealt{2003A&A...402..253S}), but hitherto known examples are typically
  much longer, tens of days. Also, in those stars the difference frequencies
  seem linked to disk feeding, for which no sign was found in \mbox{HD}186567,
  except during a few short episodes in the second half of the observational
  time base.}

The frequency group $g_2$ shows a rather puzzling appearance in the wavelet
analysis. Dividing the data into three sub-sets and analyzing them
individually by LS reveals an ensemble of frequencies that is possibly not
entirely identical to each other in frequency and certainly quite variable in
amplitude, but far more stable than one would expect from a mere time scale,
as are the \v{S}tefl frequencies of \mbox{StH$\alpha$}166 (both shown in
Fig.~\ref{fig:quarters}). Also that $g_2$ in \mbox{HD}186567 has a rather
sharp border in frequency space (see Fig.~\ref{fig:KIC011per}) is not really
what one would expect from aperiodic variablity of a given time scale.

One possibility is certainly that the sub-frequencies of $g_2$ as apparent in
Fig.~\ref{fig:quarters} are real, and the pattern seen in the wavelet analysis
is the result of a beat pattern in which the variable amplitudes of the
individual sub-frequencies destroy the coherent appearance.

{ To investigate this possibility a total of 40 B stars from the CoRoT
  database, flagged as $\beta$\,Cep or SPB stars by either
  \citet{2009A&A...506..471D} or \citet{2013A&A...550A.120S}, were analyzed in
  the same way (the full results will be presented in a future work of this
  series on the Be stars found in the CoRoT database). Two $\beta$\,Cep stars,
  \object{CoRoT ID101024938} and \object{CoRoT ID106113525} showed unresolved
  strips of multi-mode pulsational power like \mbox{HD}186567. However, their
  appearance is far more regular, that is power minima and maxima follow a
  clearly ordered and periodic pattern both in temporal sequence and in
  frequency position.

 Therefore, since the frequencies were found to be stable in
 Fig.~\ref{fig:quarters}, the distinct appearance in Fig.~\ref{fig:KIC011per}
 must be due to unstable amplitudes. Looking for possible explanations,} the
wavelet analysis may also be compared to what \citet{2009Sci...324.1540B} have
suggested to be the signature of stochastic oscillations for
{ the $\beta$ Cepheid} \object{V1449\,Aql} (see their Fig.~2). Stochastically excited oscillations
are theoretically predicted to occur across the B star spectral range
\citep[see,
  e.g.,][]{2013MNRAS.430.1736S,2014A&A...565A..47M,2014MNRAS.443.1515L}. Judging
from Fig.~6 { of \citeauthor{2013MNRAS.430.1736S}}, however, the observed
amplitude of 1\,mmag seems to be about 1 order of magnitude higher than what
is expected. Given the fast rotation of a Be star, the observed frequency
being twice as high as the highest one in the co-rotating frame of the star is
less of a concern, as will be seen below. For the case of V1449\,Aql this
explanation is not unequivocally accepted; \citet{2011A&A...534A..98A} have
suggested non-linear resonance arising from the interaction of a large number
of other modes, while \citet{2013MNRAS.431.2554D} proposes that the dominant
radial mode of V1449\,Aql produces this signature through chaotic behavior. In
any case, the similarity of the signature in the wavelet analysis remains
striking, and the positive amplitude correlation between $A_2$ and $A_3$ could
well be a clue towards the model of \citeauthor{2011A&A...534A..98A}

{\begin{table}[b!]
\caption{\label{tab:KIC06_f2} Frequencies belonging to the $g_2$ group of
  \mbox{StH$\alpha$}166, found in the first 120 steps of the LS analysis. The epoch for the
  phase is $\mathrm{BKJD}=0.0$.}
\begin{center}\small
\begin{tabular}{ccc}
Freq.   & Phase & Semi-ampl.\\
{[c/d]} &   [$0\dots1$]    & [mmag]\\
\hline
  1.027900   &     0.135   &   8.24\\
  1.047964   &     0.977   &   3.41\\
  1.026608   &     0.531   &   1.74\\
  1.029056   &     0.764   &   1.43\\
  1.048985   &     0.722   &   1.03\\
  1.007836   &     0.584   &   0.85\\
  1.046672   &     0.454   &   0.82\\
  1.032661   &     0.280   &   0.80\\
  1.008856   &     0.378   &   0.67\\
\end{tabular}
\end{center}

\end{table}}

\subsection{Rotation and frequencies in Be stars} \label{sec:rot}

For non-radial pulsations in a rotating star, the observed period is not the
one in the co-rotating frame of the star. While in most stars the rotation
rates are low enough for the difference to be of only minor importance, this
is not so for Be stars. In general, the observed and co-rotating frequencies
$f_\mathrm{obs}$ and $f_\mathrm{corot}$ are related to each other via
\begin{equation}
f_\mathrm{obs} = f_\mathrm{corot} - m \Omega,
\end{equation}
where $\Omega$ is the rotation frequency of the star and $m$ is the azimuthal
number of the non-radial pulsation. By convention negative $m$ denote a
pulsation that is propagating prograde in the co-rotating frame of the
star. Obviously, a prograde mode cannot be observed at frequencies lower than
$\Omega$. For retrograde modes the relation between observed and co-rotating
frequency can be less obvious. A mathematically negative $f_\mathrm{obs}$
means the variation appears prograde to the observer, but
$|f_\mathrm{obs}|>\Omega$ is well possible. For rapid rotators, as Be stars
are, all the above means that the simple classification as $p$- or $g$-mode
pulsator, based solely on the value of $f_\mathrm{obs}$ being lower or higher
than some division line (often 4\,c/d is chosen as delimiter), becomes
invalid. For example, a perfectly ordinary $g$-mode, say
$f_\mathrm{corot}=$1\,c/d, with $\ell=2,m=-2$, on a perfectly ordinary Be
star, say rotating at 1.5\,c/d, would be observed at 4\,c/d, and hence be
classified as a $p$-mode. For this reason, stars like \mbox{ALS}10705 or
\mbox{HD}186567 might not actually be hybrid pulsators, as they were
classified, but pure $g$-mode pulsators.


Rotation also poses a certain challenge to the suggestion by
\citet{2015MNRAS.450.3015K} that combination frequencies are responsible for
non-linear appearance of some light-curves. Combination frequencies are
observed frequencies that relate to other observed frequencies via
\begin{equation}
f_{n_i,n_j,\dots} = n_i f_i + n_j f_j + \cdots.
\end{equation}
However, in order to interact in such a way as to produce non-linear effects,
as suggested by \citeauthor{2015MNRAS.450.3015K}, the interaction must be
physical in the object, as otherwise the variations would just superimpose
linearly, and not produce detectable power in the periodogram at the
combination frequency.  In other words, the simple combination with integer
numbers $n$ must exist also in the co-rotating frame of the star. This only
works if the sum of all terms $ n_i m_i $ ($m$ again the azimuthal number),
becomes zero, which severely limits the potential of combination frequencies
to explain frequency groups, at least for rapid rotators.

Indeed, while for the slowly rotating $\gamma$\,Dor star KIC8113425
\citeauthor{2015MNRAS.450.3015K} only need four genuine frequencies to explain
39 further frequencies as combinations, the ratio is the opposite for the Be
star \mbox{HD}186567: For 15 frequencies they regard as genuine (all of which
are from the $g_2$ group, which is discussed in some detail above), they only
find four combination frequencies present in the periodogram.

\section{Conclusions}\label{sec:conclusion}
Among the stars observed by the Kepler spacecraft, three have been identified
as Be stars in the literature. By good fortune, these three provide good
examples covering the full range of Be star behavior: \mbox{ALS}10705 is
inactive without disk; \mbox{HD}186567 exhibits increasing disk feeding by
outbursts, possibly with a decaying previous disk; and \mbox{StH$\alpha$}166
is a fully active Be star showing strong, ongoing outbursts. The commonalities
and differences found are suggested to be a general scheme for the photometric
behavior of Be stars as follows:\smallskip
 
\noindent
{\bf The stellar power spectrum of a Be star in an inactive phase.}\smallskip
 
\noindent
Comparing the results for the Be star \mbox{ALS}10705, observed in quiescence,
to the literature, nothing is found in the LS analysis that is different in
principle from the appearance of non-emission B stars. That Be stars have a
tendency to be classified as hybrid pulsators, even when well outside the
$\mathrm{\beta}$~Cep instability strip, is possibly a consequence of their
rapid rotation, rather than due to the true presence of high-order
$p$-modes. In the co-rotating frame of the star all photospheric periods can
well reside in the low frequency $g$-mode regime, depending on their mode
properties $\ell$ and in particular $m$.

There are correlations between the amplitude variability of individual modes (or
mode groups, if understood as a beating phenomenon) which need an
explanation. In case of single modes that might be resonant mode-coupling in
the stellar interior, in the case of mode groups a mechanism would have to be
found; none has to our knowledge been proposed. Similar correlations for SPB
and $\mathrm{\beta}$~Cep stars are not known, but that could rather be due to
  the usage of LS dominating the literature instead of wavelet analysis for
  such stars.

Any LS analysis over the entire time span of observations should, therefore,
only be the first step in analyzing Be star variability; it must be followed
by time-resolved ways to analyze the variation, such as wavelet, or LS with a
running box-filter in the time domain. In this work the wavelet approach was
adopted for its robustness \citep[see discussion in][]{1996AJ....112.1709F}
and the frequency resolution being a function of frequency of the form of
$R=\Delta f/f=\mathrm{const.}$, which is well suited for the problem of
tracking the behavior of frequency groups over a wide range of frequencies.

A variable amplitude is also seen in the $g_2$ group of \mbox{HD}186567,
without being resolved into beatings even in the full Kepler run.  From an
observational point of view, amplitude variability on timescales longer than
the time base of typical observing runs is the norm in Be stars, not the
exception, regardless of whether one interpretes this variability as intrinsic
or as due to unresolved beating.\smallskip

\noindent
{\bf The circumstellar power spectrum of a Be star in an active phase, but
  without outbursts.} \smallskip

\noindent
For the first about 1000 days of the Kepler observations, \mbox{HD}186567 did
not show any sign of out bursts, which began around $t=1000$ and continued
until the end of the observation.  However, already before the outbursts
started, its power spectrum differed from that of the fully inactive
\mbox{ALS}10705 by the additional presence of broad bumps in the power
spectrum. Above, these were associated with the presence of a disk in the
circumstellar environment, and as \mbox{HD}186567 turned to disk feeding by
outbursts they increased in strength and moved towards longer frequencies. The
same behavior of the bumps can be seen in \mbox{StH$\alpha$}166 comparing the
times of strong and weak disk feeding.

In summary, it seems that the bumps are associated with the plain presence of
a disk, and are stronger when the disk is actively fed. In terms of the viscous
decretion disk model, in which the evolution of the disk after its formation is
fully governed by viscous processes \citep[see][for a
  review]{2013A&ARv..21...69R}, this would mean that the bump strength and
position depends on the density and possibly density slope of the innermost
parts of the disk.\smallskip

\noindent
{\bf The circumstellar power spectrum of a Be star in an active phase with
  outbursts.} \smallskip

\noindent
When a Be star is having clear outbursts, it shows the broad bumps as
discussed above, but stronger, in addition to the purely stellar
variations. Whether the bumps are associated to the outbursts themselves,
that is the ejection of matter, or the subsequent re-accretion of (part of) the
ejected matter cannot be said with the data at hand, both remains possible. It
should be noted that such re-accretion is an unavoidable feature of the
viscous processes governing the disk.

Whether some part of the purely stellar variations are attenuated, as seems to
be the case in some stars observed by CoRoT
\citep[e.g.,][]{2009A&A...506...95H}, can also not be said with the Kepler data
alone. In any case the frequency spectrum, in terms of well defined groups,
seems poorly populated compared to \mbox{ALS}10705 and \mbox{HD}186567. The
literature shows that the possibility of attenuation in strong outburst phases
exists, see also \citet{2003A&A...411..167S} for spectroscopically equivalent
behavior of $\omega$\,CMa, and the absence of the spectroscopically found
frequencies of $\mu$\,Cen in BRITE data (see Paper I).


In addition to the above, the power spectrum increases at a characteristic
frequency range, typically in the range 1--2\,c/d at a value some 10\% lower
than that of a strong stellar pulsation mode. This latter, stellar pulsation,
could be spectroscopically identified as $\ell=m=2$ $g$-modes in most stars
\citep{2003A&A...411..229R}.

For the transient frequencies in outbursts the term \v{S}tefl frequencies is
used, following Paper I. \v{S}tefl frequencies are immediately related to
outbursts, that is the transfer of matter (and angular momentum) from the star
into the circumstellar disk. Any individual event is short lived, and
subsequent events do not repeat at the same frequency or with any phase
coherency. However, as a phenomenon they are well defined and have been
identified in a number of stars (see discussion in Paper I).

\smallskip

In further works of this series, the above findings will be tested against the
databases collected by the MOST \citep{2003PASP..115.1023W}, CoRoT
\citep{2009A&A...506..411A}, and BRITE Constellation satellites
\citep{2014PASP..126..573W}, and possibly the Kepler K2 mission. Data
available from MOST, CoRoT, and K2 databases are of similar precision as the
Kepler data used here, but provide shorter time bases for analysis. However,
the Be stars in these databases are much more numerous, offering the
opportunity to get a statistical hold on the above described phenomena. BRITE
data, in turn, are less precise, and the typical campaign lengths are similar
to MOST, CoRoT, or K2. However, BRITE focuses on stars with $V<5$\,mag,
meaning it observes targets for which potentially a lot of archival data exist
and often considerably detailed analysis works have been published.  BRITE has
the potential to link the variability types found in the Kepler data to other
observables from a wide range of techniques, and will allow to make the
connection to the actual physical processes much more robust than is possible
with Kepler, MOST, K2, or CoRoT data alone. Paper I of this series, presenting
the analysis of the BRITE observations of \mbox{$\mathrm{\mu}$}~Cen and \mbox{$\mathrm{\eta}$}~Cen, serves as an
example for the power of such an approach.

\begin{acknowledgements}
The authors dedicate this work to Stanislav (Stan) \v{S}tefl (1955--2014) in
whose honor the \v{S}tefl frequencies are named.\smallskip\\

A.~C.~C acknowledges support from CNPq (grant 307594/2015-7) and FAPESP (grant
2015/17967-7).
The authors are very grateful to an anonymous referee for comments serving to
clarify the results.
This paper includes data collected by the Kepler mission. Funding for the
Kepler mission is provided by the NASA Science Mission directorate.
Some of the data presented in this paper were obtained from the Mikulski
Archive for Space Telescopes (MAST). STScI is operated by the Association of
Universities for Research in Astronomy, Inc., under NASA contract
NAS5-26555. Support for MAST for non-HST data is provided by the NASA Office
of Space Science via grant NNX09AF08G and by other grants and contracts.
This research used the facilities of the Canadian Astronomy Data Centre
operated by the National Research Council of Canada with the support of the
Canadian Space Agency.
Partly based on observations obtained at the Canada-France-Hawaii Telescope
(CFHT) which is operated by the National Research Council of Canada, the
Institut National des Sciences de l'Univers of the Centre National de la
Recherche Scientique of France, and the University of Hawaii.
This research has made use of NASA's Astrophysics Data System Service, as well
as of the SIMBAD database, operated at CDS, Strasbourg, France.

\end{acknowledgements}

\bibliographystyle{bibtex/aa} \bibliography{bibtex/sismo}

\end{document}